# Finding Convex Hulls Using Quickhull on the GPU

Stanley Tzeng and John D. Owens
University of California, Davis

**Abstract**

We present a convex hull algorithm that is accelerated on commodity graphics hardware. We analyze and identify the hurdles of writing a recursive divide and conquer algorithm on the GPU and divise a framework for representing this class of problems. Our framework transforms the recursive splitting step into a permutation step that is well-suited for graphics hardware. Our convex hull algorithm of choice is Quickhull. Our parallel Quickhull implementation (for both 2D and 3D cases) achieves an order of magnitude speedup over standard computational geometry libraries.

## 1 Introduction

Convex hull finding algorithms can be viewed as a data-parallel problem. For each iteration, the operations done per point is the same. Each point goes through some test to see whether or not it should be discarded, and those points which are left are regrouped and tested again until the final set of convex hull points are determinted. Given the popularity of convex hulls in mathematical and graphical applications along with its data-parallel nature, it would be ideal if convex hull finding could be done on manycore machines.

The current trend of Graphics Processing Unit (GPU) computing is to move intensive data-parallel computation off the CPU and onto the GPU. The programmability and peak computation power of GPUs make them a worthy candidate to transfer convex hulls onto. For some applications, a typical workflow involving convex hulls would involve producing a set of points on the CPU or (increasingly these days) on the GPU; computing the convex hull on the GPU; and then rendering (or further processing), with only the points on the convex hull, on the GPU. Computing the convex hull on the GPU would save expensive communication between CPU and GPU. The fastest methods of finding a convex hull find it in $O(n \log n)$ time using recursive divide and conquer methods where $n$ is the number of points in $d$-dimensional space. Unfortunately, current GPU architectures do not support recursion directly.

In this paper we present a convex hull algorithm on commodity graphics hardware. Our method gives rise to a generalized framework for divide and conquer algorithms on the GPU which can be used for other divide and conquer applications. Our method is based on permuting the data in the input array rather than splitting them into separate arrays. The focus of our work is the Quickhull algorithm [4]. Quickhull shares many similarities with Quicksort [8] but is more complex because of the dimensionality of the input data. Another difference is that Quickhull also throws away points from the input data it has deemed cannot be on the hull. Furthermore, Quickhull's complexity demonstrates the generality of our approach. To maintain organization after the permutation, we use the concept of *segments* [5]. Our contribution is a convex hull algorithms and a framework that makes GPUs a suitable platform for divide and conquer algorithms.

The rest of the paper is organized as follows: Section 2 goes over previous work for convex hulls and divide and conquer frameworks on the GPU. Section 3 provides an overview of divide and conquer frameworks and how it can be mapped onto the GPU along with an overview of the Quickhull algorithm. Section 4 shows the implementation of crucial operations for our framework and the final Quickhull algorithm on the GPU. In the end of this section, we describe an extension to compute convex hulls in four dimensions and higher using our framework. Section 5 shows the results of our Quickhull against popular de-facto standards of convex hull finders. Finally we conclude with future work and some examples of how our framework can be used to express other divide and conquer algorithms on manycore systems.

## 2 Previous Work

**Convex Hulls** Earliest convex hull algorithms ran in 2D planar space and had a time complexity of $O(nh)$ [9, 16]. One of the earliest $O(n \log n)$ algorithms was by Graham and used divide and conquer sorting methods to achieve the lower bound run time [11]. Divide and conquer convex hull algorithms began with the work of Preparata and Hong [21] and early implementations of the Quickhull algorithm focused on the planar case [6, 10, 12]. Barber et al. extended it to higher dimensions [4]. There have also been randomized algorithms using half-space projections which also have an expected running time of $O(n \log n)$ [9]. Parallelizing a convex routine started with using a batch of microprocessors to process the point-in-triangle tests and passing the boolean results through a series of gates [2]. These methods were later formalized into parallel languages 20 years later but these implementations were done on MIMD machines as opposed to SIMD [19].

**Divide and Conquer** A wide survey of divide and conquer algorithms can be found in the seminal textbook by Cormen et al. [8]. Hardwick defines the term *nested data parallelism* to describe the parallel subclass of divide and conquer algorithms [14]. Nested data parallel algorithm exhibit both data-parallel (one operation across multiple data) and control parallel (one function across multiple processors). Early work in parallelizing divide and conquer algorithms came as extensions to then-popular parallel programming languages for multiprocessor systems across clusters [13, 18]. Many well known divide and conquer applications have been parallelized using these cluster programming models. Perhaps the closest related to our work is the work done by Blelloch for parallel systems [5] (including 2D quickhull) and later moved onto GPUs by Sengupta et al. [22].

## 3 Theory

In this section we go over a generalized framework of divide and conquer algorithms and justify the approach on our framework. We conclude with a review of the original Quickhull algorithm.

### 3.1 Overview of Divide and Conquer

All divide and conquer algorithms share the similar step of partitioning data into smaller subsets and operating on the subsets via recursive calls. Depending on the algorithm, some can also discard data during each recursive call, cutting down on the workload as the algorithm goes deeper into the recursion. We then need to capture the following behavior: a divide step that splits up the input, and a discard step that throws away parts of the original input.

A naive way to capture this would be the straightforward approach of splitting the input array into multiple sub-arrays and process each sub-array in parallel. While this approach may work, it would be highly inefficient for two reasons. Firstly, it will suffer from load balancing issues because the amount of work needed for each sub-array will vary due to the varying number of elements in each sub-array [14]. Second, splitting the original input array into multiple sub-arrays makes for inefficient memory accesses on the GPU.

Our framework works by grouping elements within the original array into *segments*. Segments are contiguous partitions of the input data which are maintained by segment flags [5, 22]. A set flag marks the beginning of a segment (called the segment *head*). Operations in one segment do not affect elements in another segment. Because our kernels run on the original array we do not subdivide and thus avoid the load balancing issue. Thus we replace the divide step with a "permute and segment" step which regroups data and updates segment flags to let kernels know the length of each segment.

## 3.2 Overview of Quickhull Algorithm

In Algorithm 1 we show the Quickhull algorithm, which takes as input a set of points in $d$ dimensions. As an example, let us consider the 2D case. Quickhull begins by finding both the minimum and maximum (extrema) points along the $x$ dimension and adding them to the convex hull. The input set of points is divided into two subsets: those that are below the line joining the extrema points and those above. For each subset, the distance of each point to the dividing line is computed. The farthest point thus found is added to the convex hull. Then a triangle is constructed with the farthest point and the dividing line. All points in the subset which lies in the interior of this triangle are discarded. This step of division is carried on recursively with each subset until no points remain in any subset.

---

**procedure** $d$ Dimensional Quickhull
 1: {**First Split**}
 2: Find an $d$ dimensional face with extrema points. Add these points to the convex hull.
 3: Divide the remaining points into two sets: above and below the face.
 4: For each point, find distance from point to face.
 5: {**Recursive Step**}
 6: **for** each set **do**
 7:   Pick the point with max distance. Add these points to the convex hull.
 8:   Form a simplex with the points and the face.
 9:   Test each point for inclusion within the simplex. If it is inside, then it cannot be part of the convex hull. Throw the point out.
10:   For each point, find the face of the simplex closest to it. Split into $n$ sets.
11:   For each point, compute the distance from the point to the face.
12: Repeat until all points have been processed.

**Algorithm 1:** $d$-dimensional Quickhull Pseudocode.

---

## 4 Implementation

We define two operations to capture the functionality of divide and conquer algorithms: `flagPermute` which allows us to permute elements and create new segments, and `compact` which allows us to re-

```
[ 1 0 0 1 0 1 0 0 ] # s
[ 0 0 0 3 0 5 0 0 ] # s_t
[ 0 0 0 3 3 5 5 5 ] # s_h
[ 2 0 1 1 1 2 2 1 ] # f
[ 0 1 0 0 0 0 0 0 ] # mask[0]
[ 0 0 1 0 0 0 0 0 ] # scan[0]
[ 1 1 1 0 0 0 0 0 ] # backscan[0]
[ 0 0 1 1 1 0 0 1 ] # mask[1]
[ 0 0 0 0 1 0 0 0 ] # scan[1]
[ 0 0 0 1 1 0 0 0 ] # backscan[1]
[ 1 0 0 0 0 1 1 0 ] # mask[2]
[ 0 1 1 0 0 0 1 2 ] # scan[2]
[ 2 0 1 3 4 6 7 5 ] # p
[ 1 1 1 1 0 1 1 0 ] # s'
```

Figure 1: An example of how a `flagPermute` with three states (0,1, and 2) works. This is the output from the algorithm from Algorithm 2.

move elements from a segment, and removing a segment altogether when all elements in a segment are removed.

Our implementation is based on the parallel prefix sum operations found in the CUDA Parallel Primitives Library [15], and so we will be using various scan terminology to describe our operations.

### 4.1 flagPermute

`flagPermute` is our permutation routine that groups data with the same flag into segments and inserts segment flags to mark the boundaries of these new segments. The inputs and outputs to `flagPermute` are:

- **Input:** An array of state flags $f$.
- **Input:** An array of segment heads $s$.
- **Input:** The number of elements $n$.
- **Output:** An array of permutation locations $p$.
- **Output:** A new segment array $s'$.

Each element must know two things to calculate its final position after permutation:

1. How many total elements of other states are present in the segment?
2. How many elements of its own state are present to its left in the segment?

`flagPermute` finds both with a series of masks and scans. The mask outputs 1 for all elements with the state we are calculating final positions for, and 0 otherwise. Executing a forward scan gives the final position for the state we chose. It also gives us the total number of elements with that state. We do a backward max scan to distribute this total number over all other elements in the segment. This number is used to calculate indices for elements with other states.

To accomplish our goal of "permute and segment", `flagPermute` updates segment flags so that each group of states becomes a new segment. `flagPermute` reuses the information needed to find $p$ to compute the locations to insert segment heads. Algorithm 2 gives the pseudocode of the `flagPermute` routine and Figures 1 and 2(a) gives a pseudocode and visual example respectively of a`flagPermute` with three states. Belloch's `split-and-segment` [5] can be seen as a special case of `flagPermute` with two states.

## 4.2 `compact`

`compact` is our routine that removes elements within a segment. The inputs and outputs to `compact` are:

- **Input:** A boolean array $b$.
- **Input:** A segment array $s$.
- **Input:** The number of elements $n$.
- **Output:** A permutation array $p$ which contains the locations for all elements in $b$ marked as true.
- **Output:** A permutation array $sp$ which contains the permutation locations for segment heads.

Our implementation of compact works on the same principle as stream compaction for GPUs [22] but we also update or remove segment heads. At the time of compaction, one of two following scenarios can happen to any segment within the whole array:

1. *There is at least one true element which may not be the head of the segment.* We reposition the segment head to the first true element to the right. The segment head is untouched if the first element is true.
2. *All elements in a segment are marked false.* We remove the segment by discarding all data and removing the segment head.

Listing 3 gives the pseudocode of the `compact` routine and Figure 2(b) is a visual example of `compact`.

## 4.3 Quickhull

**Initialization**  Our implementation of Quickhull on CUDA [20] (which we will call CUDAQuickhull) takes as arguments:

- **Input:** An array of $n$ points $p\_in$.
- **Output:** An array of points $ch\_out$ which contains the points in $p\_in$ that make up the convex hull.

At the beginning, our segment array $s$ is initialized to 0 everywhere except for the first element which is set to 1. At the end of the algorithm, $p\_in$ will be empty since we will be either throwing away elements or moving them into $ch\_out$. We implement the algorithm described in Section 3.2 in CUDA. Listing 4 is an algorithmic overview of CUDAQuickhull in 2D.

**procedure** flagPermute($f, s, n$) **returns** $p$
1: **for** $id = 0$ to $n - 1$ **do** {in parallel}
2:   **if** $s[id] = 1$ **then**
3:     $s\_t[id] \Leftarrow id$
4:   **else**
5:     $s\_t[id] \Leftarrow 0$
6: {set scan to forward inclusive sum scan}
7: $s\_h \Leftarrow$ segmented_scan($s\_t$, $s$)
8: **for** $i = 0$ to $k - 2$ **do**
9:   maskArray[$i$] $\Leftarrow$ mask($f$, $i$)
10:   {set scan to forward exclusive sum scan}
11:   scanArray[$i$] $\Leftarrow$ segmented_scan(maskArray[$i$],$s$)
12:   {set scan to backward inclusive max scan}
13:   backscanArray[$i$] $\Leftarrow$ segmented_scan(scanArray[$i$],$s$)
14: maskArray[$k-1$] $\Leftarrow$ mask($f$, $k-1$)
15: {set scan to forward exclusive sum scan}
16: scanArray[$k-1$] $\Leftarrow$ segmented_scan(maskArray[$k-1$],$s$)
17: {`size` returns the size of an array}
18: $p \Leftarrow$ flagPermuteKernel(scanArray, backscanArray, $s\_h$, $f$, $n$)
19: {Now compute the new segment heads}
20: $s \Leftarrow$ addSegKernel(backscanArray, $s$, $s\_h$, $n$)
**end procedure** flagPermute

**function** flagPermuteKernel(scan, bs, $s\_h$, $f$, $n$) **returns** $p$
1: {flagPermuteKernel takes the scan data collected for all $k$ states and computes the final permutation location for each index $id$.}
2: **for** $id = 0$ to $n - 1$ **do** {in parallel}
3:   loc $\Leftarrow 0$
4:   {$f[id] - 1$ goes through all previous states before $f[id]$}
5:   **for** $i = 0$ to $f[id] - 1$ **do**
6:     loc $\Leftarrow$ loc + bs[$i$][$id$]
7:   $p \Leftarrow$ loc $+ s\_h[id] +$ scan[$i$][$id$]
**end function** flagPermuteKernel

**function** addSegKernel(backscan, $s$, $s\_h$, $k$ $n$) **returns** $s$
1: {addSegKernel uses the backwards scan information to figure out where to place the new segment flags.}
2: **for** $id = 0$ to $n - 1$ **do** {in parallel}
3:   **for** $i = 0$ to $nk - 1$ **do**
4:     **if** backscan[$i$][$id$] + $s\_h[id] = id$ **then**
5:       $s[id] = 1$
**end function** addSegKernel

**Algorithm 2:** $k$-state `flagPermute`. In the pseudocode scan is the scan operation and segmented_scan is the segmented scan operation defined by Belloch [5].

**procedure** compact(*b*, *s*, *n*) **returns** *p*, *sp*
1: {set scan to forward exclusive sum scan}
2: $p \Leftarrow \text{scan}(b)$
3: $n \Leftarrow p[n-1] + b[n-1]$ {this updates the number of true elements}
4: {now we update the segment}
5: $sp \Leftarrow \text{segmentMaskKernel}(b, p, n)$
6: {set scan to backward inclusive min scan}
7: $sp \Leftarrow \text{segmented\_scan}(\text{newSegment})$
**end procedure** compact

**function** segmentMaskKernel(*b*, *p*, *n*) **returns** *output*
1: **for** $id = 0$ to $n-1$ **do** {in parallel}
2:   **if** bVal is 1 **then**
3:     $output[id] \Leftarrow p[id]$
4:   **else**
5:     $output[id] \Leftarrow$ identity of min operator in scan
**end function** segmentMaskKernel

**Algorithm 3:** Pseudocode for the `compact` routine. In the pseudocode scan is the scan operation and segmented_scan is the segmented scan operation defined by Belloch [5].

**procedure** 2DCUDAQuickhull (*p_in*) **returns** *ch_out*
1: Construct *s*. All elements 0 except $s[0] = 1$.
2: {**First Split**}
3: Use min and max segmented scan to find the extrema along x axis. Add the extrema points to *ch_out* and remove from *p_in* via `compact`.
4: Test each point against the line formed by the extrema points and assign a flag depending on whether the point lies above or below the line.
5: Use 2-state `flagPermute` to rearrange *p_in* into two segments: one with points above and another below the line.
6: {**Recursive Step**}
7: **repeat**
8:   **for** each segment **do**
9:     Find the farthest point from the line.
10:     Add point to *ch_out*.
11:     Form triangle with line and point.
12:     Triangle inclusion test for each point. Those that are inside are discarded with `compact`.
13:     Assign each point a state flag depending on which edge of triangle it is closest to.
14:     Call 2-state `flagPermute` to split the segment based on the state flags.
15:     {Thus all points closest to an edge are now in one segment}
16: **until** *p_in* is empty

**Algorithm 4:** 2D CUDAQuickhull

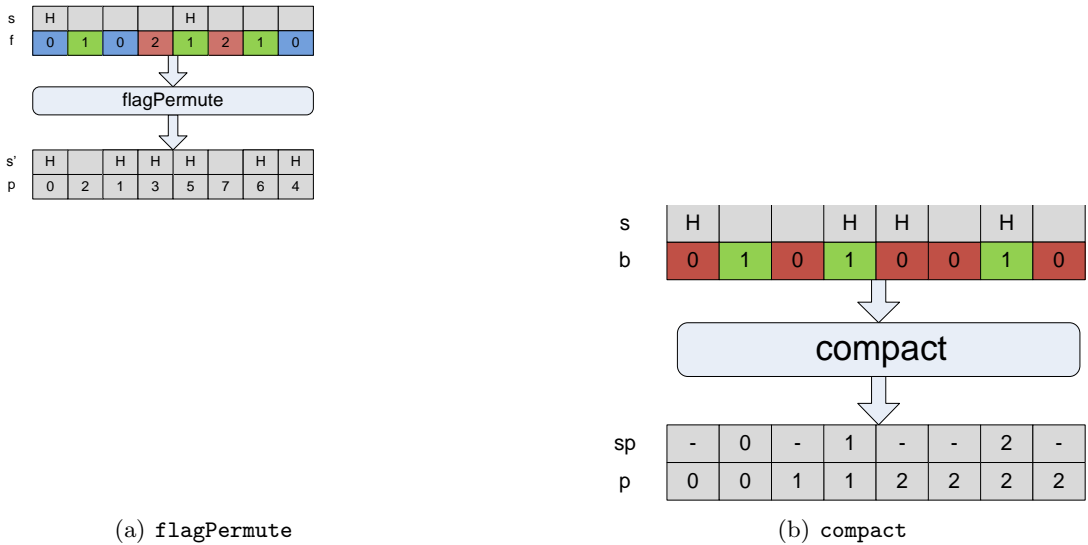

(a) `flagPermute`  (b) `compact`

Figure 2: **Left:** `flagPermute`, the input of segments heads $s$, flags $f$ where the colors of the flags denote their state. The output is a new segment array with the permutation locations for each element of the flag array. **Right**: `compact`, a boolean array $b$ and a segment array $s$, compact computes the permutation array $p$ which consists of the permute locations of the true elements and $sp$, a permutation array for the segment heads. Notice that in the third segment, both elements are false, so the segment head flag is lost. `-` indicates the identity for the min operator.

**3D and Higher Dimensions** We extend our two dimensional Quickhull algorithm (Listing 4) to three dimensions. In the First Split stage, points are now divided by a plane. The plane is formed by a triangle. One side of the triangle is the line connecting two extrema points along the x direction. The third point is the point in the input which is furthest from the line. In the Recursion Step, a point can now be associated to 3 new faces of the tetrahedra and thus we use a 3-state `flagPermute` to ensure that the points are separated correctly.

Just as the 3D case is an extension of the 2D case, the 4D case is an extension of the 3D case. A 4D simplex would generate 4 new faces per recursive call so we would have a 4-way split per segment. This would call for a `flagPermute` with four states. Since `compact` is always based on a boolean case, we do not need to modify the operation for higher-dimensional usage. The amount of memory required to store all the intermediate data will most likely be a limiting case.

## 5 Results

In this section we show the results of our GPU-based Quickhull algorithm and compare it to other popular convex hull implementations. We compared against the popular CPU-based `qhull` library [1] for convex hulls and Clarkson's CPU-based `hull` program [7]. Our test platform was an NVIDIA GTX 260 GPU with 896 MB of RAM and CUDA 2.2 and an Intel Core 2 Quad Q6600 CPU. In all tests, data starts and finishes in the memory of the processor on which the algorithm

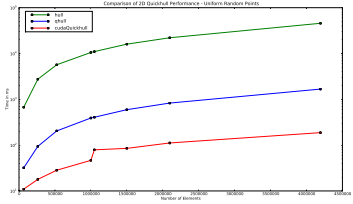

(a) 2D Uniform Points

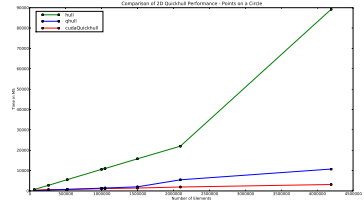

(b) 2D Points on Circle

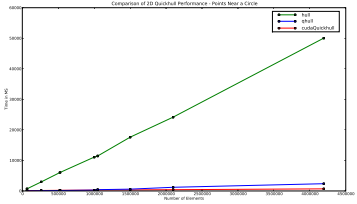

(c) 2D Points Almost on Circle

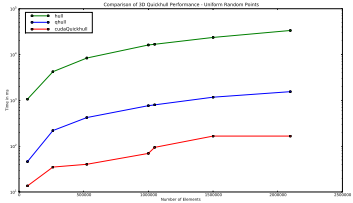

(d) 3D Uniform Points

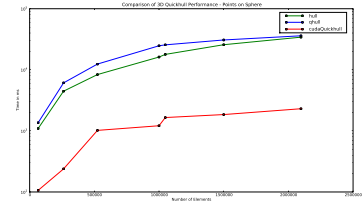

(e) 3D Points on Sphere

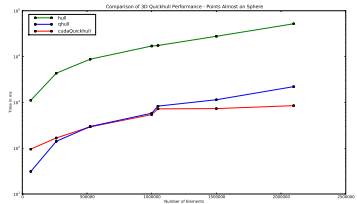

(f) 3D Points Almost on Sphere

Figure 3: Timing results for various point distributions. Our CUDAQuickhull outperforms the CPU-based convex hull algorithms for each trial.

is run, as we expect would be the most common usage.

For 2D and 3D points, we measure the time taken to build convex hulls in the following scenarios:

- $n$ points distributed randomly and uniformly within the unit circle/sphere;
- $n$ points distributed on the unit circle/sphere (thus, each point is on the convex hull); and
- $n$ points distributed near the unit circle/sphere.

From the figures in Figure 3, we can see that our GPU algorithm outperforms both CPU algorithms by about an order of magnitude. However, GPUs take a much larger hit in slowdown when each point is part of the convex hull (circle and sphere for 2D and 3D respectively). The circle case is 17 times slower than the uniform distribution case. The sphere is 14 times slower than the uniform distribution case. For the CPU cases, `qhull` has a slowdown of 23 times for the sphere case, and 6 times slowdown for the circle case when compared with the uniform distribution case. `hull` has a 2 times slowdown for the circle case and no slowdown for the circle case when compared with the uniform distribution case. The GPU slowdown is most likely due to the cost of each iteration on the GPU version. Each iteration is expensive since it involves a new set of kernel launches and a GPU to CPU memory transfer to notify the CPU of the new input size (the number of unprocessed points). For the circle and sphere cases, Quickhull must run more iterations to find the whole convex hull. Current graphic hardware only allows kernels to be launched from the CPU, so all control flow decisions require data to be transferred from the GPU back to the CPU. We expect future hardware revisions that give control flow decisions to the GPU, such as the ability to launch kernels from kernels (not available yet), will greatly reduce this overhead.

## 6    Conclusion

In this paper we have demonstrated how a divide and conquer convex hull algorithm can be implemented efficiently on commodity graphics hardware. Our results were on the order of one magnitude greater in terms of speed than those of the CPU versions. However, we expect this gap to grow as newer generations of graphics hardware is released. Applications of a fast convex hull algorithm will allow real time collision detection, recognizing textured objects, and pattern recognition [3, 17, 23].

With the growing popularity of many core computing, there is a need for algorithmic framework to explain a rich set of algorithms. In this paper we adapt a divide and conquer framework to graphic processors by framing divides as permutations and recursion as segmented operations. Our framework provides us with a language to express various other divide and conquer algorithms on the GPU. Belloch's parallel Quickhull algorithm, for example, can be expressed as a 2-state `flagPermute` without any `compact` operations. One algorithm that would use both a 2-state `flagPermute` and a `compact` operation would be $k$-median finding in linear time. For median finding, one can think of the initial scan stage as a 2-state `flagPermute` and then a `compact` stage to throw away the segment that does not contain $k$. Future work in this area includes finding CPU-GPU switchover points (i.e. after how many elements does the GPU overhead make sense?). Also, it would be interesting to see how future graphics hardware and languages will change our framework.